\documentclass[a4paper,12pt]{article}
\usepackage{amssymb}
\newcommand{\Integer}{\mathbb Z}

\newfont{\bg}{cmr10 scaled\magstep3}
\newcommand{\Bzero}{\smash{\lower1ex\hbox{\bg 0}}}
\newcommand{\bra}[1]{\langle\,#1\,|}
\newcommand{\ket}[1]{|\,#1\,\rangle}

\newcommand{\nn}{\nonumber\\}
\newcommand{\calT}{\mathcal{T}}
\newcommand{\dsp}{\displaystyle}

\begin{document}
\title{A regularization of field theory\\
on non-commutative torus}
\author{{\sc Naofumi Kitsunezaki}\thanks{Email address:
kitsune@eken.phys.nagoya-u.ac.jp}~~and
{\sc Shozo Uehara}\thanks{Email address:
uehara@eken.phys.nagoya-u.ac.jp}\\
{\it Department of Physics, Nagoya University,}\\
{\it Chikusa-ku, Nagoya 464-8602, Japan}}
\date{}
\maketitle
\vspace{-75mm}
\begin{flushright}
	DPNU-03-24\\
	hep-th/0310127\\
	October 2003
\end{flushright}
\vspace{45mm}
\begin{abstract}
Matrix model is used as a regularization of field theory on
non-commutative torus.
However, there exists an example that the product of the large-$N$
limit of matrices does not coincide with that of the corresponding
fields.
We propose a new procedure for regularizing fields on a
non-commutative torus by matrices with the help of the projection in
the representation space, so that the products of the matrices
coincide with those of the corresponding fields in the large-$N$
limit.
\end{abstract}

\section{Introduction}
It is known that the low energy effective theory with constant
background $B_{\mu\nu}$ fields on a D-brane is a field theory in the
non-commutative spacetime \cite{JHEP9909032}.
And field theory in the non-commutative spacetime, which we call
non-commutative field theory (NCFT), has been investigated in view of
studying properties of superstring theories.

It is reported that some open Wilson lines are gauge invariant in
non-commutative spacetime in Yang-Mills theory \cite{JHEP0005023}.
The stringy properties of NCFT were discovered in \cite{JHEP0002020}
using perturbative approach, which is very similar to matrix
theories.
Indeed, it was shown that real fields on non-commutative discrete
periodic lattice (NCDPL) of $N\times N$ sites can be mapped to
$N\times N$ Hermitian matrices, and their integration over the lattice
are mapped to the trace of the corresponding matrices
\cite{PRD62105018}.
To construct the basis of fields over NCDPL of $N\times N$ sites, let
us consider
\begin{equation}
 T^{(N)}_{(m,n)}(\hat{x},\hat{y})= e^{2i\kappa(m\hat{x}+n\hat{y})},
	\label{eq:PL}
\end{equation}
where $\kappa$ is a constant which will be determined later,
$[\hat{x},\hat{y}]=i$ and $m$ and $n$ are integers, which makes the
$(\hat{x},\hat{y})$-space periodic,
\begin{equation}
 T^{(N)}_{(m,n)}(\hat{x}+k\alpha,\hat{y}+l\alpha)
	= T^{(N)}_{(m,n)}(\hat{x},\hat{y})\,.
	\qquad (k,l\in\Integer,\alpha=\pi/\kappa)
\end{equation}
In order to regularize by matrix we will impose periodicity on $m$ and
$n$ to restrict the representation space of eq.(\ref{eq:PL}) by the
condition
\begin{equation}
  e^{2i\kappa N\hat{x}}=e^{2i\kappa N\hat{y}}=1.
\end{equation}
Actually this leads to $T^{(N)}_{(m+kN,n+lN)}(\hat{x},\hat{y})
=T^{(N)}_{(m,n)}(\hat{x},\hat{y})$ and also this means that the
$(\hat{x},\hat{y})$-space is now discrete. Furthermore, $\kappa$ is
set $\sqrt{\pi M/N}$ where $M$ and $N$ are mutually prime since
$e^{2i\kappa N\hat{x}}$ should commute with any
$T^{(N)}_{(m,n)}(\hat{x},\hat{y})$, so that the commutation relation
becomes
\begin{equation}
 [\,T^{(N)}_{(m_1,m_2)}(\hat x,\hat y),
	T^{(N)}_{(n_1,n_2)}(\hat x,\hat y)\,]
  =-2i\sin\left(2\pi\frac{M}{N} (m_1n_2-m_2n_1)\right)\,
	T^{(N)}_{(m_1+n_1,m_2+n_2)}(\hat x,\hat y).
\end{equation}
Then, the matrix basis corresponding to eq.(\ref{eq:PL}) are
written by the clock and the shift matrices,
\begin{equation}
 M^{(N)}_{(m,n)}= e^{i2\pi\frac{M}{N}mn}
    \left(\begin{array}{@{\,}ccccc@{\,}}
	1&&&&\\
	&e^{4\pi i\frac{M}{N}}&\multicolumn{3}{c}\Bzero\\[1ex]
	&&\ddots&&\\[1ex]
	\multicolumn{3}{c}\Bzero &&e^{{4\pi i\frac{M}{N}}(N-1)}\\[1ex]
    \end{array}\right)^m
    \left(\begin{array}{@{\,}ccccc@{\,}}
	0&1&&&\\
	&0&1&\multicolumn{2}{c}\Bzero\\
	&&\ddots&\ddots&\\
	&\Bzero&&0&1\\
	1&&&&0\end{array}\right)^n.  \label{eq:CandS}
\end{equation}
Due to this correspondence, the real fields on NCDPL are mapped to
Hermitian matrices, the star products of fields are mapped to products
of the corresponding matrices and the integration over NCDPL coincide
with the trace of the corresponding matrix \cite{PRD62105018}.
These facts suggest that NCFT on torus can be regularized by the
corresponding matrix model.

On the other hand, we find that the matrices in eq.(\ref{eq:CandS})
can be expressed by the following operators on the Fock space,
\begin{equation}
 {\calT}^{(N)}_{(m,n)}
  =\sum_{k=0}^{N-n-1}e^{i2\pi\frac{M}{N}(2km+mn)}\ket{k}\bra{k+n}
  +\sum_{k=0}^{n-1}e^{i2\pi\frac{M}{N}(2km-mn)}\ket{N-n-k}\bra{k},
  \label{eq:CandSinF}
\end{equation}
where
\begin{equation}
  \ket{k}=\frac{1}{\sqrt{k!}}\,(\hat{a}^\dagger)^k\ket{0}\,,
	\qquad(\,\hat{a}\ket{0}=0\,)
\end{equation}
and
\begin{equation}
  \hat{a}=\frac{1}{\sqrt{2}}\,(\hat{x}+i\hat{y}),\quad
  \hat{a}^\dagger=\frac{1}{\sqrt{2}}\,(\hat{x}-i\hat{y}).
 \label{eq:AAdag}
\end{equation}
Since $T^{(N)}_{(m,n)}(\hat{x},\hat{y})$ in eq.(\ref{eq:PL}) and
${\calT}^{(N)}_{(m,n)}$ in eq.(\ref{eq:CandSinF}) are isomorphic, we
can expect that their large-$N$ limits will also be isomorphic, so
that NCFT on a torus will be completely reproduced by the large-$N$
limit of the matrix models.
However, there exists an counterexample:
In the large-$N$ limit, the operators $T^{(N)}_{(0,\pm 1)}$
have the following weak convergence limits,\footnote{We
cannot discuss the convergence of the operators with the ``norm''
which is naturally defined by the trace of the operator, since such a
``norm'' is divergent for some operators.}
\begin{eqnarray}
\lim_{N\to\infty}\calT^{(N)}_{(0,1)}~(\equiv \calT^{(\infty)}_{(0,1)})
	&=&\sum\limits_{k=0}^{\infty}\ket{k}\bra{k+1},\nn
\lim_{N\to\infty}\calT^{(N)}_{(0,-1)}~(\equiv \calT^{(\infty)}_{(0,-1)})
	&=&\sum\limits_{k=0}^{\infty}\ket{k+1}\bra{k},
\end{eqnarray}
and their product is calculated as
\begin{equation}
 \calT^{(\infty)}_{(0,-1)}\,\calT^{(\infty)}_{(0,1)}=1-\ket{0}\bra{0}\,.
  \label{eq:ng-prod}
\end{equation}
On the other hand, the product of the corresponding plane waves
in eq.(\ref{eq:PL}) are
\begin{equation}
   T^{(\infty)}_{(0,-1)}(\hat{x},\hat{y})\,
	T^{(\infty)}_{(0,1)}(\hat{x},\hat{y})=1,\label{eq:Id}
\end{equation}
which obviously differs from eq.(\ref{eq:ng-prod}).
This raises a question that the matrix models might not regularize
NCFT.

In this paper, we consider the following projected plane-wave like
(PPWL) operators,
\begin{equation}
 {\widetilde \calT}^{(N)}_{(m,n)}=\sum_{k,l=0}^{N-1}
  \ket{k}\bra{k}e^{i2\pi\lambda(m\hat a+in{\hat a}^\dagger)}
  \ket{l}\bra{l},  \label{eq:prj}
\end{equation}
where $\lambda$ is an arbitrary real number and
$\hat{a},\hat{a}^\dagger$ are given in eq.(\ref{eq:AAdag}), and
hence all the operators of $\exp[2\pi i\lambda(m\hat{a}
+in{\hat{a}}^\dagger)]$ are equivalent to the plane waves
$\exp[2\pi i\lambda(m\hat{x}+n\hat{y})]$.\footnote{They are equivalent
but not $unitary$ equivalent.} And we show that $N^2$ numbers of
PPWL operators made of the lowest $N$ states in
eq.(\ref{eq:prj}) can be used as the bases of $N\times N$
Hermitian matrices instead of the clock and the shift matrices in
eq.(\ref{eq:CandSinF}).  The advantage of using PPWL operators are
that the products of the projected plane waves in
eq.(\ref{eq:prj}) are equal to those of the original plane waves
even in the large-$N$ weak convergence limit,
\begin{eqnarray}
&&\lim\limits_{N\to\infty}
  \Biggl|\,\sum\limits_{p=0}^{N-1}
   \bra{k}e^{i2\pi\lambda(m_1\hat a+im_2{\hat a}^\dagger)}\ket{p}
   \bra{p}e^{i2\pi\lambda(n_1\hat a+in_2{\hat a}^\dagger)}\ket{l}\nn
&&\qquad\qquad   - \bra{k}\,
   e^{i2\pi\lambda(m_1\hat a+im_2{\hat a}^\dagger)}\,
   e^{i2\pi\lambda(n_1\hat a+in_2{\hat a}^\dagger)}\,
   \ket{l}\,\Biggr| =0\,. \label{eq:eval}
\end{eqnarray}
In eq.(\ref{eq:eval}), we have used the fact that the projection
operator $P_N=\sum_{p=0}^{N-1}\ket{p}\bra{p}$ becomes an identity
operator in the large-$N$ weak convergence limit, so that there is no
such problem as the discrepancy between eqs.(\ref{eq:ng-prod}) and
(\ref{eq:Id}).
Then the question to ask is whether $N^2$ numbers of PPWL operators
are independent or not, which is shown in the next section.
The final section is devoted to summary and discussion.

\section{Independence of PPWL operators}
In this section, we show that any $N^2$ numbers of the PPWL operators
in eq.(\ref{eq:prj}) are independent, i.e.,\
$\sum\limits_{m,n}a_{(m,n)}{\widetilde\calT}^{(N)}_{(m,n)}=0$ leads to
$a_{(m,n)} =0$\ for all $(m,n)$.
Let us define $N^2\times N^2$ matrices $X^{(N)}=
\Bigl(X^{(N)}_{(m,n)(k,\,l)}\Bigr)$ as
\begin{eqnarray}
 X^{(N)}_{(m,n)(k,\,l)}&\equiv&
	\bra{k}e^{i2\pi\lambda(m\hat a+in{\hat a}^\dagger)}\ket{l}
	 \label{eq:define-X}\\
 &=&e^{-\pi^2\lambda^2(m^2+n^2)}\sqrt{k!\,l!}
  \sum_{p=0}^{\min\{k,l\}}\frac{(i\sqrt{2}\pi\lambda m)^{k-p}
  (-\sqrt{2}\pi\lambda n)^{l-p}}{p!(k-p)!(l-p)!},
	 \label{eq:M-eles}
\end{eqnarray}
where $(m,n)$ run over some different $N^2$ points in $\Integer^2$
and $0\le k,l\le N-1$.
Then the above statement is equivalent to the one that the determinant
of the $N^2\times N^2$ matrix $X^{(N)}$ is non-zero.
Since the factor $e^{-\pi^2\lambda^2(m^2+n^2)}$ in
eq.(\ref{eq:M-eles}) depends only on the number of the rows of the
matrix $X^{(N)}$ while $\sqrt{k!\,l!}$ on the number of the columns,
we have
\begin{equation}
 \det X^{(N)}
  =\left(\prod_{m,n}e^{-\pi^2\lambda^2(m^2+n^2)}\right)
	\left(\prod_{k,l=0}^{N-1}\frac{1}{\sqrt{k!\,l!}}\right)\,
	\det {\widetilde X}^{(N)},
\end{equation}
where
\begin{eqnarray}
 {\widetilde X}^{(N)}_{(m,n)(k,\,l)} &=&
    k!\,l!\sum_{p=0}^{\min\{k,l\}}
    \frac{(i\sqrt{2}\pi\lambda m)^{k-p}\,
	(-\sqrt{2}\pi\lambda n)^{l-p}}{p!\,(k-p)!\,(l-p)!}\,,\\
 &=& (i\sqrt{2}\pi\lambda m)^k(-\sqrt{2}\pi\lambda n)^l
  +kl\,(i\sqrt{2}\pi\lambda m)^{k-1}(-\sqrt{2}\pi\lambda n)^{l-1}\nn
 &&\qquad+\frac{k(k-1)\,l(l-1)}{2}\,(i\sqrt{2}\pi\lambda m)^{k-2}
	(-\sqrt{2}\pi\lambda n)^{l-2}+\cdots,\label{eq:tldX2}
\end{eqnarray}
so that we have only to show $\det {\widetilde X}^{(N)}\ne0$.
It is still difficult to directly calculate the determinant.
Notice that the second term in eq.(\ref{eq:tldX2}) is
proportional to the first term of the corresponding expansion of
${\widetilde X}^{N}_{(m,n)(k-1,\,l-1)}$,
the third term in eq.(\ref{eq:tldX2}) is proportional to the second
term of the expansion of ${\widetilde X}^{N}_{(m,n)(k-1,\,l-1)}$ which
is also proportional to the first term of the expansion of
${\widetilde X}^{N}_{(m,n)(k-2,\,l-2)}$ and so on. Then we can write
\begin{equation}
 {\widetilde X}^{(N)}_{(m,n)(k,\,l)} =
  (i\sqrt{2}\pi\lambda m)^k(-\sqrt{2}\pi\lambda n)^l
  +kl {\widetilde X}^{(N)}_{(m,n)(k-1,\,l-1)}+\cdots.\label{eq:tldX3}
\end{equation}
It is obvious that the ellipsis in eq.(\ref{eq:tldX3}) can be
rewritten by a proper linear combination of the elements
$\{{\widetilde X^{(N)}_{(m,n)(k-2,\,l-2)},\cdots,
{\widetilde X^{(N)}}_{(m,n)(k-t,\,l-t)}}\}$ where $t=\mbox{min}\{k,l\}$,
i.e., we find that there exist such coefficients
$\{a^{(k,\,l)}_p\}_{p=1}^{\min\{k,\,l\}}$ that the matrix element
${\widetilde X}^{(N)}_{(m,n)(k,\,l)}$ can be written by
\begin{equation}
 {\widetilde X}^{(N)}_{(m,n)(k,\,l)}
  =(i\sqrt{2}\lambda m)^k(-\sqrt{2}\lambda n)^l
  +\sum_{p=1}^{\min\{k,\,l\}}
	a^{(k,\,l)}_p{\widetilde X}^{(N)}_{(m,n)(k-p,\,l-p)}\,.
\end{equation}
Thus, due to the fundamental property of the determinant,\footnote{The
coefficients $a^{(k,\,l)}_p$ do not depend on $m$ and $n$.} we have
\begin{equation}
 \det {\widetilde X}^{(N)}=\det Y^{(N)},
\end{equation}
where the matrix elements of
$Y^{(N)}=\left(Y^{(N)}_{(m,n)(k,l)}\right)$ is given by
\begin{equation}
 Y^{(N)}_{(m,n)(k,\,l)}=(i\sqrt{2}\lambda m)^k(-\sqrt{2}\lambda n)^l,
\end{equation}
which is given by the direct product of the Vandermonde matrices, and
hence the determinant of $Y^{(N)}$ is non-zero.
Thus we have shown that any $N^2$ numbers of PPWL operators in
eq.(\ref{eq:prj}) made of the lowest $N$ states are
independent. Especially, if we take the set of operators
$\{\widetilde{\calT}^{(N)}_{(m,n)}: -(N-1)/2\le m, n\le(N-1)/2\}$,
which are of course independent, we can easily take a proper linear
combination to make Hermitian elements, so that we can use them as the
bases of  $N\times N$ Hermitian matrices.

\section{Summary and discussion}
We have studied how field theory on the non-commutative torus is
regularized by $N\times N$ Hermitian matrix theory.
In the usual regularization with the clock and the shift matrices in
eq.(\ref{eq:CandS}), some products of the matrices do not recover the
original products of the fields even in the large-$N$ limit as we have
seen in eqs.(\ref{eq:ng-prod}) and (\ref{eq:Id}), which concerns that
matrix theory could not reproduce field theory on non-commutative
torus even in the large-$N$ limit.
Note that the large-$N$ limit here is the weak convergence limit,
which means that the matrices converge element by element.
Furthermore, notice that non-zero matrix elements
$\bra{k}T^{(N)}_{(m,n)}\ket{l}$ in eq.(\ref{eq:CandSinF}) may change
$N$ by $N$. This causes a difference between the large-$N$ limit of
the product of the matrices and the product of the large-$N$ limit of
the matrices.
On the other hand, if we adopt the PPWL operators in
eq.(\ref{eq:prj}), there appears no such problem since matrix elements
$\bra{k}{\widetilde\calT}^{(N)}_{(m,n)}\ket{l}\ (0\leq k,l\leq N-1)$
do not change  in the large-$N$ limit. Thus we can safely use the
matrix model to regularize non-commutative field theory.

Finally, we shall comment on field theory with area-preserving
diffeomorphism, which is usually argued as the large-$N$ limit of
matrix model \cite{Hoppe,deWit:1988ig,zachos}.
At finite $N$, the basis of $N\times N$ Hermitian matrices are usually
chosen by
\begin{equation}
 {\widetilde T^{(N)}}_{(m,n)}(\hat{x},\hat{y})=\frac{N}{2\pi}\,
	e^{2i\sqrt{\pi/N}(m\hat{x}+n\hat{y})},\label{eq:PL2}
\end{equation}
or
\begin{equation}
 {\widetilde M}^{(N)}_{(m,n)}=\frac{N}{2\pi}\,
  e^{i\frac{2\pi}{N}mn}
  \left(\begin{array}{@{\,}ccccc@{\,}}
	1&&&&\\
	&e^{i\frac{4\pi}{N}}&\multicolumn{3}{c}\Bzero\\[1ex]
	&&\ddots&&\\[1ex]
	\multicolumn{3}{c}\Bzero &&e^{{i\frac{4\pi}{N}}(N-1)}\\[1ex]
   \end{array}\right)^m
  \left(\begin{array}{@{\,}ccccc@{\,}}
	0&1&&&\\
	&0&1&\multicolumn{2}{c}\Bzero\\
	&&\ddots&\ddots&\\
	&\Bzero&&0&1\\
	1&&&&0
   \end{array}\right)^n,  \label{eq:CandS-2}
\end{equation}
instead of $T^{(N)}_{(m,n)}(\hat x,\hat y)$ in eq.(\ref{eq:PL}) or
$M^{(N)}_{(m,n)}$ in eq.(\ref{eq:CandS}), respectively.
In the large-$N$ limit, their commutation relations are expected to
become
\begin{equation}
  [\,{\widetilde T}^{(\infty)}_{(m_1,m_2)}(\hat{x},\hat{y}),
    {\widetilde T}^{(\infty)}_{(n_1,n_2)}(\hat{x},\hat{y})\,]
  =-2i(m_1n_2-m_2n_1)\,
    {\widetilde T}^{(\infty)}_{(m_1+n_1,m_2+n_2)}(\hat{x},\hat{y}),
  \label{eq:Poisson-T}
\end{equation}
or
\begin{equation}
  [\,{\widetilde M}^{(\infty)}_{(m_1,m_2)},
    {\widetilde M}^{(\infty)}_{(n_1,n_2)}\,]
  =-2i(m_1n_2-m_2n_1)\,
     {\widetilde M}^{(\infty)}_{(m_1+n_1,m_2+n_2)},
  \label{eq:Poisson-M}
\end{equation}
which are the same algebra as the Poisson bracket of the plane waves
on the two-dimensional torus.
As we pointed out before, changing matrix elements when one enlarges
the size of the matrix was the reason of breaking isomorphism in the
large-$N$ limit, thus we need a set of generators of the Poisson
algebra represented on the Fock space to make PPWL operators, so that
we can check the validity whether the field theory on torus with
area-preserving diffeomorphism can be regularized by matrix model
without breaking the isomorphism.

The following operators satisfy the same commutation relations as
those of the Poisson bracket of the plane waves over two-dimensional
torus \cite{KU02},
\begin{equation}
 {\widetilde{\calT}}^{(P,\xi)}_{(m_1,m_2)}
  =\left\{1-2i\sqrt{\pi}\left(\xi m_1\hat{a}
	+\frac{m_2}{\xi}\hat{a}^\dagger\right)\right\}
    \exp\left[2i\sqrt{\pi}\left(\xi m_1\hat{a}
	+\frac{m_2}{\xi}\hat{a}^\dagger\right)\right],
\end{equation}
where $\xi$ is an arbitrary real constant.
Then we would expect that a similar procedure holds. That is, once we
take the $N^2$ numbers of the operators whose indices satisfy
$-(N-1)/2\le m_1,m_2\le (N-1)/2$ and project out to the lowest $N$
states, we would have the bases of the $N\times N$ Hermitian matrices
as in the non-commutative torus case in eq.(\ref{eq:prj}), and field
theory with area-preserving diffeomorphism could also be regularized
by matrix model.  However, if we choose such a set of $N^2$ elements
which include \footnote{We consider only odd $N$ case.}
\begin{equation}
 \{{\widetilde{\calT}}^{(P,\xi)}_{(0,m)}:~
     m=\pm 1,\pm 2,\cdots,\pm \frac{N-1}{2}\}\,,\label{eq:dep-set}
\end{equation}
we find that these $N$ elements are {\it not} independent.
To prove this, we show that there exists a non-trivial set of
$\{x_n\}$ for the equation,
\begin{equation}
 \sum_{n}x_n  \left(\sum_{k,l=0}^{N-1}\ket{k}\bra{k}
   {\widetilde{\calT}}^{(P,\xi)}_{(0,n)}\ket{l}\bra{l}
   \right)=0,   \label{eq:dependence-P}
\end{equation}
where $n$ runs over $-(N-1)/2,\cdots,-1,1,\cdots,(N-1)/2$.
Since the matrix elements of (\ref{eq:dep-set}) are given by
\begin{equation}
 \bra{k}{\widetilde{\calT}}^{(P,\xi)}_{(0,n)}\ket{l}
  =\left\{\begin{array}{ll}
   \dsp\sqrt{\frac{k!}{l!}}\,\frac{1}{(k-l-2)!}
   \left(-\frac{2\sqrt{\pi}}{\xi}n\right)^{k-l}&(k\ge l+2)\\[2ex]
   {}~~0&(k=l+1)\\[.5ex]
   {}~~1&(k=l)\\[.5ex]
   {}~~0&(k<l)\,,\end{array}\right.
\end{equation}
eq.(\ref{eq:dependence-P}) reduces to the following $N-1$ equations,
\begin{equation}
 \begin{array}{ll}
  \dsp\sum_n~ x_n&=0,\\
  \dsp\sum_n\left(-\frac{2\sqrt{\pi}}{\xi}n\right)^2x_n&=0,\\
  \multicolumn{2}{c}\vdots \\
  \dsp\sum_n\left(-\frac{2\sqrt{\pi}}{\xi}n\right)^{N-1}x_n&=0.
 \end{array}\label{eq:eqn-sys}
\end{equation}
Thus, to prove the existence of non-trivial $\{x_n\}$, we have only to
show
\begin{equation}
 \det\left(\begin{array}{cccccc}
       1&\cdots&1&1&\cdots&1\\
       \left(\frac{2\sqrt{\pi}}{\xi}n_{\frac{N-1}{2}}\right)^2
	&\cdots
	&\left(\frac{2\sqrt{\pi}}{\xi}n_1\right)^2
	&\left(-\frac{2\sqrt{\pi}}{\xi}n_1\right)^2
	&\cdots
	&\left(-\frac{2\sqrt{\pi}}{\xi}n_{\frac{N-1}{2}}\right)^2\\
       \vdots& &\vdots&\vdots& &\vdots\\
       \left(\frac{2\sqrt{\pi}}{\xi}n_{\frac{N-1}{2}}\right)^{N-1}
	&\cdots
	&\left(\frac{2\sqrt{\pi}}{\xi}n_1\right)^{N-1}
	&\left(-\frac{2\sqrt{\pi}}{\xi}n_1\right)^{N-1}
	&\cdots
	&\left(-\frac{2\sqrt{\pi}}{\xi}n_{\frac{N-1}{2}}\right)^{N-1}
      \end{array}
     \right)=0. \label{eq:P-det}
\end{equation}
And the above Vandermonde-like determinant can be straightforwardly
shown to be zero by an elementary calculation.
This implies that the similar projection does not work in this case
and we should be very careful to regularize field theory with
area-preserving diffeomorphism by matrix model \cite{KU02}.
This is consistent with the fact reported in \cite{KU01,ANO},
where the $N$ behavior of the matrix model and the lattice
regularization of the corresponding field theory with area-preserving
diffeomorphism have been numerically compared.

\vspace{1em}
\noindent {\bf Acknowledgments:}\\
The work of SU is supported in part by MEXT Grant-in-Aid for the
Scientific Research \#13135212.

\end{document}